\documentclass[aps,prl,twocolumn,superscriptaddress,floatfix,nofootinbib,longbibliography]{revtex4-2}

\usepackage{graphicx}
\usepackage{dcolumn}
\usepackage{bm}
\usepackage{amsmath}
\usepackage{amsfonts}
\usepackage{amssymb}
\usepackage{physics}
\usepackage{float}
\usepackage{orcidlink}
\usepackage{hyperref}
\hypersetup{
    colorlinks = true,
    citecolor  = blue,
    linkcolor  = blue,
    urlcolor   = blue
}

\begin{document}

\title{$r$-Process Nucleosynthesis with {\it Ab Initio} Nuclear Masses around the $N=82$ Shell Closure}

\author{J.~Kuske\orcidlink{0009-0005-5121-7343}}
\email{jan.kuske@tu-darmstadt.de}
\affiliation{Technische Universit\"at Darmstadt, Department of Physics, 64289 Darmstadt, Germany}

\author{T.~Miyagi\orcidlink{0000-0002-6529-4164}}
\email{miyagi@nucl.ph.tsukuba.ac.jp}
\affiliation{Center for Computational Sciences, University of Tsukuba, 1-1-1 Tennodai, Tsukuba 305-8577, Japan}
\affiliation{Technische Universit\"at Darmstadt, Department of Physics, 64289 Darmstadt, Germany}
\affiliation{ExtreMe Matter Institute EMMI, GSI Helmholtzzentrum f\"ur Schwerionenforschung GmbH, 64291 Darmstadt, Germany}
\affiliation{Max-Planck-Institut f\"ur Kernphysik, Saupfercheckweg 1, 69117 Heidelberg, Germany}

\author{A.~Arcones\orcidlink{0000-0002-6995-3032}} 
\email{almudena.arcones@tu-darmstadt.de}
\affiliation{Technische Universit\"at Darmstadt, Department of Physics, 64289 Darmstadt, Germany}
\affiliation{GSI Helmholtzzentrum f\"ur Schwerionenforschung GmbH, 64291 Darmstadt, Germany}
\affiliation{Max-Planck-Institut f\"ur Kernphysik, Saupfercheckweg 1, 69117 Heidelberg, Germany}

\author{A.~Schwenk\orcidlink{0000-0001-8027-4076}}
\email{schwenk@physik.tu-darmstadt.de}
\affiliation{Technische Universit\"at Darmstadt, Department of Physics, 64289 Darmstadt, Germany}
\affiliation{ExtreMe Matter Institute EMMI, GSI Helmholtzzentrum f\"ur Schwerionenforschung GmbH, 64291 Darmstadt, Germany}
\affiliation{Max-Planck-Institut f\"ur Kernphysik, Saupfercheckweg 1, 69117 Heidelberg, Germany}

\begin{abstract}
Our understanding of the origin of heavy elements beyond iron relies on the rapid neutron-capture process ($r$-process), which accounts for roughly half of their cosmic abundance. However, the extreme neutron-rich conditions required for the $r$-process involve many nuclei that remain experimentally inaccessible, making theoretical predictions essential. We explore the impact of nuclear masses calculated with the {\it ab initio} valence-space in-medium similarity renormalization group, focusing on the region around the $N=82$ shell closure. We show for the first time that such {\it ab initio} mass calculations can be used to refine $r$-process predictions compared to global, but more phenomenological mass models. With the {\it ab initio} masses, the waiting point of the second $r$-process peak is strengthened, which leads to an overall slower nucleosynthesis flow, lower abundances of nuclei beyond the peak, and a stronger shift of the third $r$-process peak.
\end{abstract}

\maketitle

\emph{Introduction.--} Half of the heavy elements in the Universe are produced by the $r$-process where neutron captures on freshly synthesized seed nuclei are rapid compared to their beta decays. This nucleosynthesis process runs along extremely neutron-rich nuclei, for which experimental information is scarce. Therefore, theoretical models have to be used, which strongly disagree away from stability \cite{Arnould2007_RprocessStellarNucleosynthesis, Cowan2021_OriginHeaviestElements}. Our understanding of the origin of heavy elements in the Universe has improved dramatically in recent years with the observation of gravitational waves from the merger of two neutron stars and its associated electromagnetic counterparts \cite{Abbott2017_GW170817ObservationGravitational, Drout2017_LightCurvesNeutron, Kilpatrick2017_ElectromagneticEvidenceThat}. Significant progress has been made in astrophysics simulations and observations of old stars in our galaxy and neighboring dwarf galaxies. From the nuclear physics perspective, experiments will reach more neutron-rich nuclei and theoretical efforts have advanced {\it ab initio} calculation to medium-mass and heavy nuclei. In this work, we combine {\it ab initio} calculations with a global mass model to improve nucleosynthesis calculations based on neutron star merger simulations and other astrophysical environments.

Nucleosynthesis calculations for the $r$-process require knowledge of the astrophysical conditions (i.e., evolution of the density and temperature and the initial fraction of electrons or protons over all baryons) and of the nuclear properties of the involved nuclei. These properties include the rates of neutron captures ($n,\gamma$), photodissociations ($\gamma,n$), and beta decays, as well as fission barriers and yield distributions. Therefore, nuclear masses are key as they enter in the calculation of all these inputs and have a significant impact on the reaction rates. Moreover, during the $r$-process an ($n,\gamma$)-($\gamma,n$) equilibrium is reached, during which the nucleosynthesis path is given by the neutron separation energy.

The impact of nuclear masses on the $r$-process has been extensively investigated \cite{Kratz1993_IsotopicRProcessAbundances, Kratz1998_NuclearstructureInputRprocess, Arcones2011_DynamicalRprocessStudies, Mumpower2015_ImpactIndividualNuclear, Martin2016_ImpactNuclearMass, Vassh2021_MarkovChainMonte}. Many studies are based on global mass models for all unmeasured nuclei up to the neutron drip line. Within these models the finite-range liquid drop model (FRDM) \cite{Moller2019_NuclearPropertiesAstrophysical} is extensively used and it is our reference model here. In addition to studies based on different global theoretical models, new experimental results trigger investigations where only few new measured nuclei are included and the impact of their masses is explored (see, e.g., \cite{Baruah2008_MassMeasurementsMajor, Orford2018_PrecisionMassMeasurements, Vilen2020_ExploringMassSurface, Li2022_FirstApplicationMass, Zhou2023_MassMeasurementsShow}).

Over the last two decades, the range of applicability of nuclear {\it ab initio} calculations has expanded rapidly. Combining developments in systematically improvable chiral effective field theory and quantum many-body methods, meaningful calculations for heavy nuclei are now possible (see, e.g., \cite{Hu2022, Miyagi2024, Door2025}). We are thus in a position to explore the impact of systematically computed nuclear masses on $r$-process simulations. In this work, we employ the valence-space in-medium similarity renormalization group (VS-IMSRG) \cite{Tsukiyama:2010rj,Hergert:2015awm,Stroberg2017,Stroberg:2019mxo}, one of the {\it ab initio} many-body methods applicable to ground and excited states of closed- and open-shell nuclei. In particular, VS-IMSRG calculations have been successful in predicting the trend of the two-neutron shell gap in $N=82$ isotones \cite{Manea2020_FirstGlimpse$N82$} and the empirical pairing gap in Sn isotopes \cite{Mollaebrahimi2025_PrecisionMassMeasurements}.

\emph{Objectives.--} Similarly to  studies  where few new experimental masses are used combined with a global theoretical mass calculation, our aim here is not to pin down the r-process path or astrophysical site, but rather to demonstrate the impact of incorporating {\it ab initio} theoretical masses across the $N=82$ shell closure. Since such state-of-the-art mass calculations are not feasible for all nuclei, we focus on the masses around the double-magic $^\mathrm{132}\mathrm{Sn}$. This region has been identified as critical for the production of the second $r$-process peak (associated to neutron number $N=82$) and beyond \cite{Kratz2005_RprocessIsotopes132Sn, Arcones2011_DynamicalRprocessStudies, Mumpower2015_ImpactUncertainNuclear}, but it still goes beyond the experimental reach. Moreover, before the third peak ($N=126$), there is a significant impact \cite{Vassh2025_PrivateCommunication}. In these extreme neutron-rich regions, because the strength of the shell closure and the evolution beyond is unknown, mass models yield different behaviors of the neutron separation energy that is critical to determine the $r$-process path \cite{Arcones2011_DynamicalRprocessStudies,  Arcones2012_NuclearCorrelationsProcess, Mumpower2016_ImpactIndividualNuclear, Martin2016_ImpactNuclearMass, Martinet2025_ImpactMassUncertainties}. In this work, we demonstrate how global models can be combined with VS-IMSRG calculations to gain new insight into the $r$-process and the origin of heavy elements. In doing so, our {\it ab initio} calculations then also point to where further experimental investigations are important.

\emph{Methods.--} All nucleosynthesis calculations are performed using the publicly available nuclear reaction network \textsc{WinNet} \cite{Reichert2023_NuclearReactionNetwork}. For every mass model all neutron-capture rates are recalculated with the  \textsc{Talys2.0} code~\cite{Koning2023_TALYSModelingNuclear} and the photodissociation rates are determined from detailed balance with consistent partition functions, similarly to \cite{Martin2016_ImpactNuclearMass}.  Charged-particle reaction rates are from~\cite{Cyburt2010_JinaReaclibDatabase}, beta-decay rates from \cite{Moller2019_NuclearPropertiesAstrophysical}, fission rates and distributions from \cite{Panov2005_CalculationsFissionRates, Panov2010_NeutroninducedAstrophysicalReaction, Moller2015_FissionBarriersEnd, Khuyagbaatar2020_SpontaneousFissionHalflives}. As our baseline mass model we take FRDM2012 \cite{Moller2017_NuclearGroundstateMasses} combined with experimental data from AME2020 \cite{Huang2021_AME2020Atomic, Wang2021_AME2020Atomic}. Our new VS-IMSRG masses and uncertainties will be incorporated in this baseline.

To test the robustness of our conclusions, we have repeated the nucleosynthesis calculations for two different sets of beta-decay rates: REACLIB \cite{Cyburt2010_JinaReaclibDatabase} (based on FRDM1995 masses) and D3C* \cite{Marketin2016_LargescaleEvaluationDecay} (based on a relativistic Hartree-Bogoliubov mass model). In addition, we repeated all rate and nucleosynthesis calculations for two different global mass models as baselines: HFB24 \cite{Goriely2013_FurtherExplorationsSkyrmeHartreeFockBogoliubov} and HFBD1M \cite{Goriely2016_GognyhartreefockbogoliubovNuclearmassModel}. While these result in significant variation of the final abundances (especially around $A=190$), the impact of the VS-IMSRG masses remains the same. Therefore, we limit our discussion here to the beta-decay rates from \cite{Moller2019_NuclearPropertiesAstrophysical} and FRDM2012 mass model~\cite{Moller2017_NuclearGroundstateMasses}.

\begin{figure}[t!]
\centering
    \includegraphics[width=0.9\linewidth]{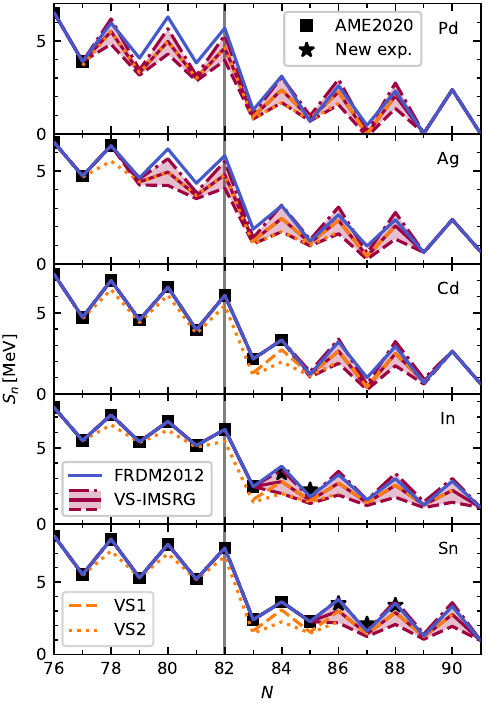}
    \caption{One-neutron separation energies $S_n$ across $N=82$ for the Pd to Sn isotopic chains, for which we explore the new VS-IMSRG masses. Our baseline FRDM2012 model (blue solid line) uses experimental masses from AME2020 (black squares) \cite{Huang2021_AME2020Atomic, Wang2021_AME2020Atomic} and otherwise the theoretical ones from \cite{Moller2017_NuclearGroundstateMasses}. Dashed and dotted orange lines are the calculated VS-IMSRG masses using two different valence spaces (VS1/VS2). Their differences for $N=83-86$ are used to estimate the red uncertainty bands using Eq.~(\ref{eq:UncertaintyConstruction}). The resulting three VS-IMSRG mass scenarios (red dashed/solid/dash-dotted lines) include the experimental masses and transition to FRDM2012 for the heaviest nuclei, beyond the VS-IMSRG calculations. The stars show recent experimental masses \cite{Izzo2021_MassMeasurementsNeutronrich, Mollaebrahimi2025_PrecisionMassMeasurements} (not included in AME2020), which are not included in the models but agree within uncertainties.}
    \label{fig:Sn_AMEmax}
\end{figure}

We have performed VS-IMSRG calculations for 70 isotopes: $^{122-134}\text{Pd}$, $^{123-135}\text{Ag}$, $^{124-137}\text{Cd}$, $^{125-139}\text{In}$, and $^{126-140}\text{Sn}$, based on nucleon-nucleon (NN) and three-nucleon (3N) interactions from chiral effective field theory.
The masses are computed with the NN+3N 1.8/2.0 (EM) interaction, which can globally reproduce ground-state energies up to $^{132}$Sn~\cite{Simonis2017,Stroberg2021, Miyagi2022_ConvergedInitioCalculations}.
The VS-IMSRG calculations start from  the Hamiltonian represented in a 15 major harmonic-oscillator space.
After normal ordering with respect to an ensemble Hartree-Fock reference~\cite{Stroberg2017}, the VS-IMSRG decouples a selected valence space from the complement. We approximate the VS-IMSRG evolution at the normal-ordered two-body level using the \texttt{IMSRG++} code~\cite{imsrg++}.
Although recent efforts make it possible to perform calculations with explicit three-body operators~\cite{Heinz2021, Stroberg2024, Heinz2025}, systematic calculations for medium-heavy nuclei are computationally still challenging. For the NN+3N interactions considered in this work, it was recently shown that this induces a $1\%$ error on the correlation energy \cite{Heinz2025}. Since ground-state energies are highly correlated, this results in even smaller corrections for separation energies.
To partly assess the many-body truncation, we employ two valence spaces: $Z=28-50$ and $N=82-126$ (VS1) and $Z=28-50$ and $N=64-90$ (VS2). The corresponding results are shown with the dashed and dotted orange lines in Fig.~\ref{fig:Sn_AMEmax}.
Within the valence space, exact diagonalizations are performed with the KSHELL code~\cite{Shimizu2019}.
We also performed calculations with the NN+3N $\Delta$N$^2$LO$_{\rm GO}$\,(394) interaction~\cite{Jiang2020}, yielding similar separation energies. Thus, for Hamiltonians applicable to medium-heavy nuclei, this dependence for $S_n$ seems smaller than the many-body uncertainty. We have also checked that VS-IMSRG(3f2) calculations~\cite{He:2024utz} with approximate triples corrections are within the VS1 and VS2 results for $S_n$.

On top of our FRDM2012 plus AME2020 baseline mass model shown in Fig.~\ref{fig:Sn_AMEmax}, we consider three mass scenarios using the new VS-IMSRG results. Starting from the most neutron-rich experimentally known mass, the VS-IMSRG separation energies are added consecutively, using VS2 (dotted) for $N \le 82$ and VS1 for $N > 82$ (dashed orange lines). Beyond the heaviest VS-IMSRG-calculated isotope, we transition back to the FRDM2012 mass model. This defines our central VS-IMSRG mass scenario, shown by the solid red line in Fig.~\ref{fig:Sn_AMEmax}.

For $N=83-86$ we have $S_n$ results from both VS1 and VS2. Their differences are used to estimate the many-body uncertainties of our calculations, because in an exact calculation both valence spaces would give identical results. Because of pairing effects, these differences are larger for even $N$. We take this into account by defining an uncertainty $\Delta S_n$
\begin{alignat}{2}
    \Delta S_n =
    \label{eq:UncertaintyConstruction}
        &\begin{cases}
        \underset{N=83,85}{\max}{|S_n^\mathrm{VS1} - S_n^\mathrm{VS2}|} \,, & \text{for odd $N$,} \\
        \underset{N=84,86}{\max}{|S_n^\mathrm{VS1} - S_n^\mathrm{VS2}|} \,, & \text{for even $N$.}
    \end{cases}
\end{alignat}
The VS-IMSRG uncertainty band in Fig.~\ref{fig:Sn_AMEmax} is obtained by adding $\pm \Delta S_n$ to our central VS-IMSRG mass scenario. We also refer to these as the min/max scenario.
The VS-IMSRG $S_n$ are generally smaller than those of FRDM2012, while the max scenario is very similar. A test of our {\it ab-initio}-based mass models is given in Fig.~\ref{fig:Sn_AMEmax} by the comparison against recently measured In~\cite{Izzo2021_MassMeasurementsNeutronrich} and Sn~\cite{Mollaebrahimi2025_PrecisionMassMeasurements} masses not included in AME2020. We find that these are consistent with our VS-IMSRG mass scenarios within estimated many-body uncertainties. Finally, we note that the amplitude of odd-even staggering in Fig.~\ref{fig:Sn_AMEmax} is increased beyond $N=86$ in Pd, Ag, and Cd isotopes, compared to the Sn and In isotopes, indicating an enhancement of the pairing gap.

\begin{figure}[t!]
\centering
    \includegraphics[width=0.9\columnwidth]{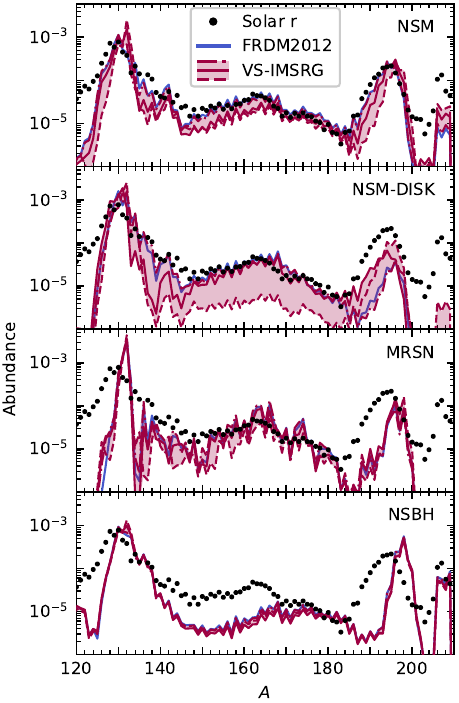}
    \caption{Final abundances for representative trajectories of a neutron star merger, its disk, a magneto-rotational supernova, and a neutron star black hole merger (see text for details). The different lines correspond to the different mass scenarios from Fig.~\ref{fig:Sn_AMEmax}. For comparison, solar $r$-process abundances~\cite{Sneden2008_NeutronCaptureElementsEarly} are shown as black dots. Note that, while we plot the different VS-IMSRG mass scenarios with a shaded band, the impact on abundances is nonlinear (as can be seen from the ordering of lines) and different masses within the $S_n$ band can give abundances outside this band.}
    \label{fig:finab_AMEmax}
\end{figure}

\begin{figure}[t!]
\centering
    \includegraphics[width=0.9\columnwidth]{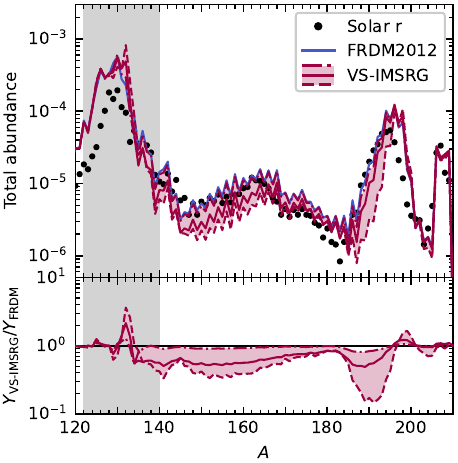}
    \caption{Top panel: Integrated abundances of 11,218 mass-weighted trajectories from \cite{Jacobi2023_EffectsNuclearMatter, Ricigliano2024_ImpactNuclearMatter} for the different mass scenarios from Fig.~\ref{fig:Sn_AMEmax}. Bottom panel: Abundance ratios of the three VS-IMSRG mass scenarios compared to FRDM2012. The gray region corresponds to the range of the new VS-IMSRG masses.}
    \label{fig:integratedFinalabundances}
\end{figure}

\emph{Nucleosynthesis results.--} We perform $r$-process calculations for trajectories from simulations of different astrophysical environments. The impact of the VS-IMSRG masses depends on the astrophysical conditions as they will determine whether and when the $r$-process path crosses the region in the nuclear chart where the {\it ab initio} calculations are included. We use individual trajectories, which are representative for the production of elements in and beyond the second $r$-process peak, from simulations of a neutron star merger \cite{Jacobi2023_EffectsNuclearMatter, Ricigliano2024_ImpactNuclearMatter} (NSM), long-time neutron star merger disk ejecta \cite{Fernandez2013_DelayedOutflowsBlack, Wu2016_ProductionEntireRange} (NSM-DISK), magneto-rotational supernovae \cite{Obergaulinger2017_ProtomagnetarBlackHole, Reichert2022_MagnetorotationalSupernovaeNucleosynthetic} (MRSN), and neutron star black hole merger \cite{Korobkin2012_AstrophysicalRobustnessNeutron, Rosswog2013_MultimessengerPictureCompact, Piran2013_ElectromagneticSignalsCompact} (NSBH). For the MRSN simulation we use a typical trajectory from the ejecta that produce an $r$-process. The final abundances for the representative trajectories are shown in Fig.~\ref{fig:finab_AMEmax} for the different mass scenarios from Fig.~\ref{fig:Sn_AMEmax}. Moreover, we demonstrate that our results are robust by calculating the mass-integrated abundances for all 11,218 trajectories of the NSM simulation (see later, Fig.~\ref{fig:integratedFinalabundances}).

The top panel of Fig.~\ref{fig:finab_AMEmax} shows the final abundances for the NSM trajectory. The abundances in the second peak at $A=132$ are higher by a factor of 2--3 when using the VS-IMSRG masses. This is due to the generally lower separation energies and increase of pairing correlations beyond $N=86$ for the Pd, Ag, and Cd isotopes. This would therefore be an exciting region to explore experimentally. Matter stops and accumulates in this region, as visible by the lower abundances beyond the second peak.

\begin{figure}[t!]
\centering
    \includegraphics[width=0.9\columnwidth]{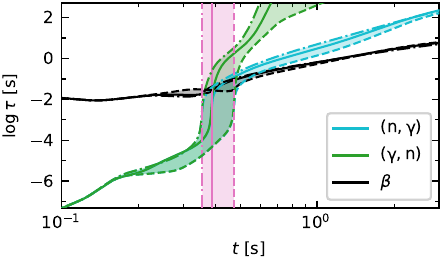}
    \caption{Averaged timescales for neutron capture ($n,\gamma$; cyan), photodissociation ($\gamma,n$; green), and beta decay ($\beta$; black) for the representative NSM trajectory (top panel of Fig.~\ref{fig:finab_AMEmax}). The different lines of the shaded band correspond to the different VS-IMSRG mass scenarios. The pink vertical lines mark the end of ($n,\gamma$)-($\gamma,n$) equilibrium and the $r$-process freeze-out.}
    \label{fig:timescales}
\end{figure}

In order to better understand the evolution and the impact of the VS-IMSRG masses, Fig.~\ref{fig:timescales} shows the averaged reaction timescales for neutron capture, photodissociation, and beta decay. Initially neutron capture and photodissociation are in equilibrium and are faster than beta decays. In this ($n,\gamma$)-($\gamma,n$) equilibrium phase, the $r$-process path runs along nuclei with a constant separation energy (see, e.g., Eq.\,(25) in \cite{Arnould2007_RprocessStellarNucleosynthesis} and Eq.\,(3) in \cite{Arcones2011_DynamicalRprocessStudies}). For the NSM trajectory the final phase of ($n,\gamma$)-($\gamma,n$) equilibrium occurs around $S_n^\mathrm{eq}\approx 1.3-1.7$\,MeV.

During the $r$-process evolution the magic number $N=82$ is reached and the path evolves along it toward larger $Z$ via beta decays and subsequent neutron captures, until this equilibrium is overcome in Rh ($Z=45$) around 0.2\,s. Now the $r$-process crosses the region where the mass scenarios differ and the evolution and reaction timescales start to depend on this: In the max VS-IMSRG scenario (dash-dotted lines; almost identical to the reference model, which is therefore not shown) the matter flow continues toward heavy nuclei afterwards. In contrast, for the central (solid) and min VS-IMSRG scenarios (dashed lines) matter stops again at $N=86$ (notice the low $S_n$ at $N=87$ for Pd, Ag, and Cd in Fig.~\ref{fig:Sn_AMEmax}). These nuclei act as a second ``waiting region'' where the flow accumulates again, before eventually beta decays\footnote{At $N=86$ the half lives are slightly longer than those of heavier nuclei and this leads to slightly higher values of the averaged beta-decay timescale for the min VS-IMSRG scenario at $t \approx 0.3$\,s.} with neutron emission move the matter toward larger proton numbers. The nucleosynthesis flow can continue toward heavier nuclei via neutron captures only once the In isotopic chain has been reached.

During this phase, neutrons are depleted, leading to the $r$-process freeze-out (vertical lines in Fig.~\ref{fig:timescales}) characterized by a fast drop in neutron density as well as a fast increase of neutron-capture and photodissociation timescales. As the evolution with the VS-IMSRG masses is slower due to the additional mass accumulation at $N=86$, the freeze-out occurs later. The last phase is the decay to stability with additional neutron captures, but negligible photodissociations. During this phase, the third $r$-process peak (see Fig.~\ref{fig:finab_AMEmax}) is shifted toward higher $A$. This is most pronounced in the min VS-IMSRG scenario because there are more neutrons left after the delayed freeze-out. Note that the interplay with beta-decay rates also emphasizes the importance of consistent {\it ab initio} calculations of beta decays \cite{Li2026_InitioCalculationsVDecay}.

The overall effect is similar for the NSM-DISK conditions, except for the central VS-IMSRG scenario, which peaks at $A=131$ instead of $132$ and for which the shift of the third peak is smallest. This reflects the intricate and non-linear interplay that nuclear masses can have on the nucleosynthesis evolution. 
For the MRSN trajectory, the temperature stays above 1\,GK for about 2.5\,s. Photodissociations keep the $r$-process path closer to stability, where beta decays are slower, making the whole evolution and the freeze-out last longer. Therefore, the region of the new VS-IMSRG masses is only marginally crossed.
The opposite trend is found for the NSBH case, where the huge neutron densities move the path far from stability beyond the region of the new VS-IMSRG masses. Therefore, in these two extreme conditions the new masses have only a minor impact. Note that for very neutron-rich conditions (i.e., NSM and NSBH), fission becomes important. Fission yields contribute to the abundances around and beyond the second peak, thus impacting the region where the masses are changed.

The discussion above is based on individual trajectories. However, this approach can overestimate the total impact of nuclear physics properties on the final abundances of an astrophysical event. Therefore, we also perform nucleosynthesis calculations for the entire mass-integrated ejecta of the NSM model of \cite{Jacobi2023_EffectsNuclearMatter, Ricigliano2024_ImpactNuclearMatter}. The results in Fig.~\ref{fig:integratedFinalabundances} show that the impact is very similar to our representative trajectory. Again, the second $r$-process waiting point is enhanced, leading to higher abundances and a later freeze-out that shifts the third peak toward larger $A$. However, in contrast to the single representative trajectory in Fig.~\ref{fig:finab_AMEmax}, the final abundances beyond the third peak are not impacted.

\emph{Conclusions.--} In this Letter, we have applied {\it ab initio} nuclear masses to $r$-process nucleosynthesis calculations. This refines the traditional approach of supplementing experimental masses with global phenomenological models. We have calculated the masses for 70 $r$-process nuclei around the $N=82$ shell closure using the VS-IMSRG method. We propagated the uncertainties for separation energies, which are dominated by many-body truncation errors, to the neutron-capture and photodissociation rates and include them in nucleosynthesis calculations for representative trajectories of different astrophysical scenarios. The VS-IMSRG masses halt the $r$-process around $N=86$ leading to a changed second peak and a later freeze-out. This produces a shift of the third peak, due to more post-freeze-out neutron captures. The same impact is found for the total ejecta of a neutron star merger simulation, based on a large set of mass-weighted trajectories, thus verifying the robustness of our results. Many $r$-process nuclei are expected to remain out of experimental reach even for the next-generation rare isotope facilities. Therefore, high-fidelity models like the VS-IMSRG including theoretical uncertainties are a promising way forward in our understanding of the $r$-process. Future studies should target a larger set of nuclei and include the impact on beta-decay half-lives and beta-delayed neutron-emission probabilities consistently.

\emph{Acknowledgments.--} We thank Jason Holt and Nicole Vassh for useful discussions. This work was supported in part by the Deutsche Forschungsgemeinschaft (DFG, German Research Foundation) -- Project-ID 279384907 -- SFB 1245, the European Research Council (ERC) under the European Union's Horizon 2020 research and innovation programme (Grant Agreement No.~101020842), the State of Hessen within the Research Cluster ELEMENTS (Project ID 500/10.006), the Japan Science and Technology Agency ERATO Grant No. JPMJER2304, and JSPS KAKENHI Grant Number~25K07294. This research in part used computational resources allocated at J\"{u}lich Supercomputing Center and provided by Multidisciplinary Cooperative Research Program in Center for Computational Sciences, University of Tsukuba.

\emph{Data availability.--} The data that support the findings of this article are openly available \citep{zenodoData}.

\bibliography{bibliography}

\end{document}